\documentclass[12pt]{iopart}
\usepackage{epsfig}
\usepackage{rotating}

\newcommand{\imag}{\mathring{\imath}}
\newcommand{\sqhz}{\ensuremath\sqrt{\mbox{Hz}}}
\newcommand{\cph}{\ensuremath\mathrm{cycles}/\sqrt{\mbox{Hz}}}
\newcommand{\mcph}{\ensuremath\mathrm{{\mu}cycles}/\sqrt{\mbox{Hz}}}

\newenvironment{idlplot}[2]{
\begin{figure}[#1]\centering
\includegraphics[width=3.5in, height=6in, angle=90]{#2}
}{\end{figure}}

\begin{document}


\article[LISA Test-bed Interferometer]{}
{Demonstration of the Zero-Crossing Phasemeter 
with a LISA Test-bed Interferometer}

\author{S E Pollack\dag\ and R T Stebbins\ddag}
\address{\dag\ JILA, University of Colorado, Boulder, CO  80309-0440\\
    {\it present address:} CENPA, Nuclear Physics Laboratory, University of Washington, Seattle, WA  98195-4290}
\address{\ddag\ NASA/GSFC Code 661, Greenbelt, MD  20771}
\ead{scott.pollack@colorado.edu}

\begin{abstract}
The Laser Interferometer Space Antenna (LISA) is being
designed to detect and study in detail gravitational waves
from sources throughout the Universe such as
massive black hole binaries.
The conceptual formulation of the LISA space-borne
gravitational wave detector is now well developed.
The interferometric measurements between the sciencecraft
remain one of the most important technological and
scientific design areas for the mission.  

Our work has concentrated on developing the
interferometric technologies to create a LISA-like
optical signal and to measure the phase of that
signal using commercially available instruments.
One of the most important goals of this research 
is to demonstrate the LISA phase timing and 
phase reconstruction for a LISA-like fringe signal, 
in the case of a high fringe rate and a low signal level.  
We present current results of a test-bed interferometer designed
to produce an optical LISA-like fringe signal previously 
discussed in \cite{jennrich} and \cite{pollack}.  
\end{abstract}


\submitto{\CQG}


\section{Introduction\label{intro}}

When a gravitational wave passes through the plane of the LISA antenna,
it can be regarded as changing the separations between the sciencecraft.
One of the key elements of the LISA mission 
is the fringe metrology system.  
The small changes in distances between sciencecraft create
small variations in the phase of the interferometric fringe formed 
at each end of each arm of the LISA interferometer.
Information from the LISA constellation will be a time series of 
phase measurements.  From this time series we will extract 
the gravitational wave signals.

There are several phasemeters under investigation for the LISA project.
Two of particular interest are the zero-crossing phasemeter
and a rapid-sampling technique.  Unlike the single-bit rapid-sampling
technique described in \cite{pollack}, 
the current popular rapid-sampling technique utilizes a multi-bit sampler.  
After digitization of the input signal the 
in-phase and out-of-phase sinusoidal
components of the signal are identified.  Taking the inverse
tangent of the quotient of these components results in the phase
of the signal.  One particular advantage of this phase-quadrature
measurement is the ability to separate positive and negative
frequencies,
hereby alleviating possible DC-wrapped noise problems (see \S\ref{noise}).

We have developed the zero-crossing phasemeter 
to show that the LISA phase measurement requirements can be
met with a simple implementation.
The zero-crossing phasemeter was described in \cite{jennrich},
along with an experiment designed to produce an optical
LISA-like fringe signal.

One of the most important milestones in any mission is verification.
By its nature, the complete operations of the LISA constellation
will not be verifiable on the ground.  However, certain aspects
of the LISA mission are ground verifiable.  
Ground testing of the LISA mission will be of key importance
for recommendation for flight by NASA and ESA.

Of particular interest to us is the interferometry measurement system.  
The experiment first described in \cite{jennrich},
and later updated in \cite{pollack} is designed to produce
the response of one end of one arm of the LISA interferometer.
Our table-top interferometer produces an interferometric fringe similar
to what LISA will produce: a slowly varying, high-frequency, low-level signal.
There are several important aspects of the LISA fringe signal
which are not included in our setup which we discuss in Section~\ref{problems}.

The techniques developed in the construction of this interferometer
will be useful for future ground-based testing of the LISA system.
Of particular interest are the modulation schemes used to simulate the 
motions of the sciencecraft and the stabilization schemes used to 
reduce laboratory disturbances.

\section{Zero-Crossing Phasemeter}

\begin{figure}[p]\centering
\includegraphics[width=\textwidth,angle=-90]{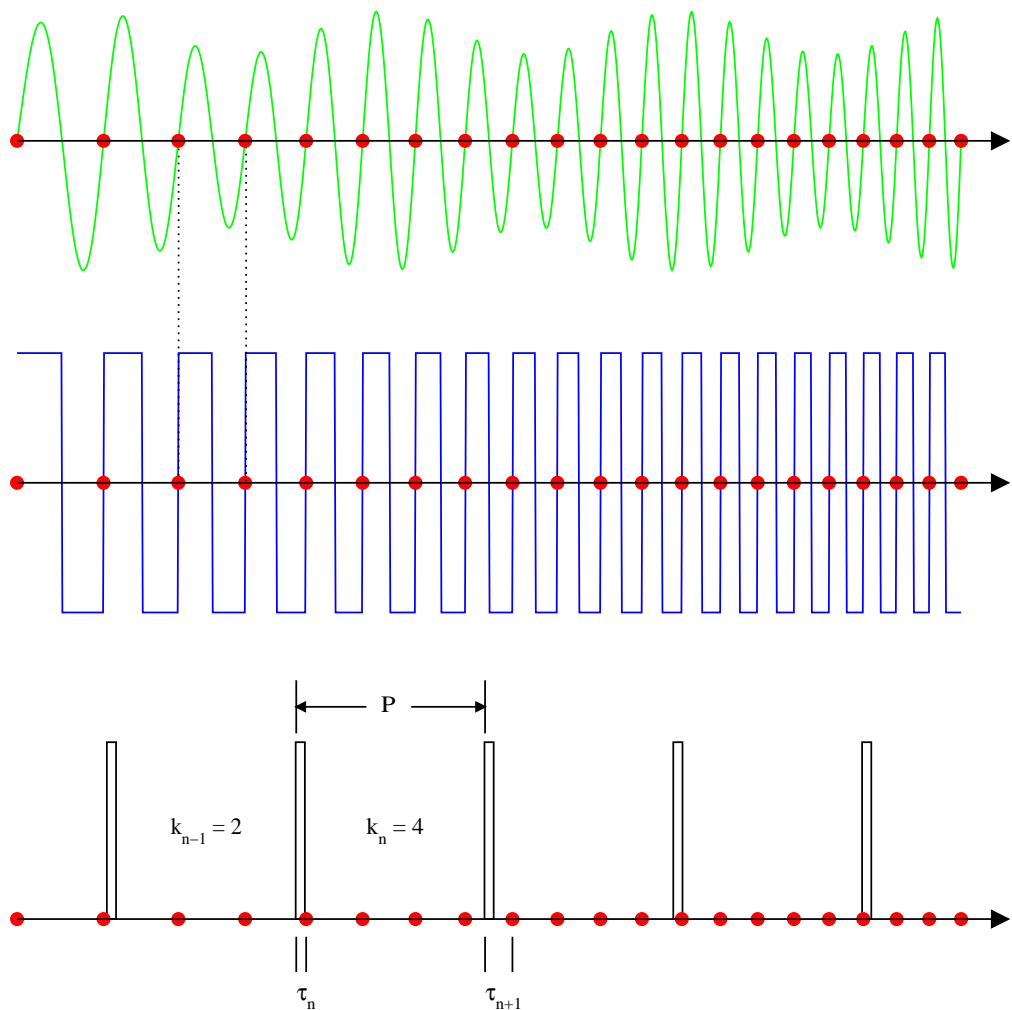}
\caption[Phase Timing Scheme]{\label{fig:timing}
The upper panel shows a typical input to our phase meter:
an amplitude varying chirping sinusoid.
The conversion of amplitude noise into
phase noise can be reduced by amplifying and clamping the initial signal.
This is depicted by the chirping square wave in the middle panel.
Amplification also makes identification of the positive going zero-crossings
(solid disks) simpler.  Each of
the vertical bars in the bottom panel
are tick marks of our fiducial clock, which
has a period $P$.  The number of zero-crossings $k_n$
between each clock tick are recorded by a counter,
and the timings $\tau_n$ between a clock tick and the first
zero-crossing are recorded by our timer.  
The equations in \cite{pollack} make use of these definitions.
}\end{figure}

The zero-crossing phasemeter which we utilize measures phase
by timing between the zero-crossings of the input signal and a fiducial clock.
We condition the signal by amplification and clipping to help
identify the zero-crossings.  
If we neglect any noise in the signal and assume a perfect phasemeter, 
then if the signal and clock frequencies are the same,
the times measured should be constant.  
This represents a
constant phase difference between the signal and the clock.
Typically the signal frequency is orders of magnitude higher
than the sampling frequency.
Figure~\ref{fig:timing} illustrates our timing scheme.

In general, we need to know the number of signal pulses during 
the interval between two timing pulses, as well as the
time difference at the beginning and end of the timing interval.
To obtain this information, we introduce a counting circuit which counts
the number of zero-crossings between clock cycles.
In this way the number of counts gives the whole number of cycles
in the timing interval to within a fraction, and a combination
of the time differences gives the correction for the fraction of a cycle.
From this collection of counts and
times we can reconstruct the phase of the signal at any point in time.
Further details on the phase reconstruction
algorithm can be found in the literature \cite{jennrich, pollack}.

Ideally the zero-crossing phasemeter can record the phase 
as a function of time for any AC signal.
Given a certain fixed timing accuracy, the measured phase noise increases 
linearly with the signal frequency.  Therefore if the signal frequency
changes by a factor of two, then the measured phase noise likewise will
change by a factor of two.  This is not an advantageous situation for LISA,
where the frequency of the fringe will range from DC to 20\,MHz.  Therefore
it would be wise to fix the signal frequency within some operating
bandwidth.  

Given the nature of the zero-crossing phasemeter, the determination
of the signal phase from the counting of zero-crossings 
alone would have 
a larger fractional error for lower frequency signals.  
However, when one includes the fractional cycle from the timing
information it becomes apparent that a lower frequency signal
results in a smaller phase error.
Nevertheless, we have chosen to make 
the signal frequency as high as possible to stay
away from DC noise sources.  
The frequency synthesizer
which we use to generate our signal has an intrinsic phase noise floor.
We have chosen the maximum frequency of our operating band to be the
cross-over frequency between the frequency dependent phase noise of
our phasemeter and the frequency independent phase noise of our
frequency synthesizer.  

\section{Zero-Crossing Phasemeter Response}

We have investigated the response of the zero-crossing phasemeter to 
various signal types using electronic input signals.  
In particular we present results
for various signal frequencies, frequency sweep rates, and
local oscillator bandwidths.

As mentioned in Section~4.1 of \cite{pollack}
the phase noise floor of the zero-crossing phasemeter, determined
by the timing resolution of the timer, is linearly proportional
to the signal frequency.  
Therefore it is advantageous to mix the input signal 
with a local oscillator (LO) to a 
suitable intermediate frequency (IF) for counting. 
At low enough frequencies the phase
noise level is dominated by intrinsic phase noise in our
electronics (as seen in Figure~3 of \cite{pollack}).

The phase noise that we present is computed by subtracting
the known phase of our signal generator from the
reconstruction of the phase from our phasemeter (see \cite{pollack}).
From this set of residuals we compute an amplitude spectral density,
the \emph{phase noise}, as a function of frequency from the signal
frequency.

\subsection{Sweep Rate Investigation}

The orbits of the LISA sciencecraft currently under investigation are such
that the distances between the sciencecraft change over the course
of a year in a roughly sinusoidal manner \cite{merkowitz, sweetser, sphughes}.  
The changing distance causes a Doppler shift between the frequencies of the
incoming and outgoing laser beams used to produce the LISA fringe
signal.  Since the distances between the orbits changes over time,
the Doppler frequency will vary over the course of a year.  The sweep
rate is the rate of change of the Doppler frequency.

The exact orbits of the LISA sciencecraft have not been determined.  
This leaves open the question of how fast the sweep rate might be.  
However, it is likely that the sweep rate will not exceed a few Hz/s.
We have performed a small investigation
mapping out the phasemeter response to signals with different sweep
rates.  Figure~\ref{fig:rate} contains data of sweeping 1\,MHz signals
mixed down to about 1\,kHz with various sweep rates.  
When the intermediate frequency (IF) 
is about to leave the 1---2\,kHz band, we
instruct the LO to step in frequency by 1\,kHz.
In this way the IF always remains between 1 and 2\,kHz.
The spectra shown are for signals with sweep rates of 
0.1\,Hz/s, 1.0\,Hz/s, and 10\,Hz/s.  For reference,
the spectrum from a constant signal frequency is shown.
We investigated sweep rates as low as 0.01\,Hz/s and as high as 1\,kHz/s.  
For the sweep rates investigated, 
the phase noise level does not differ
from the case of a constant signal frequency.

\begin{idlplot}{!p}{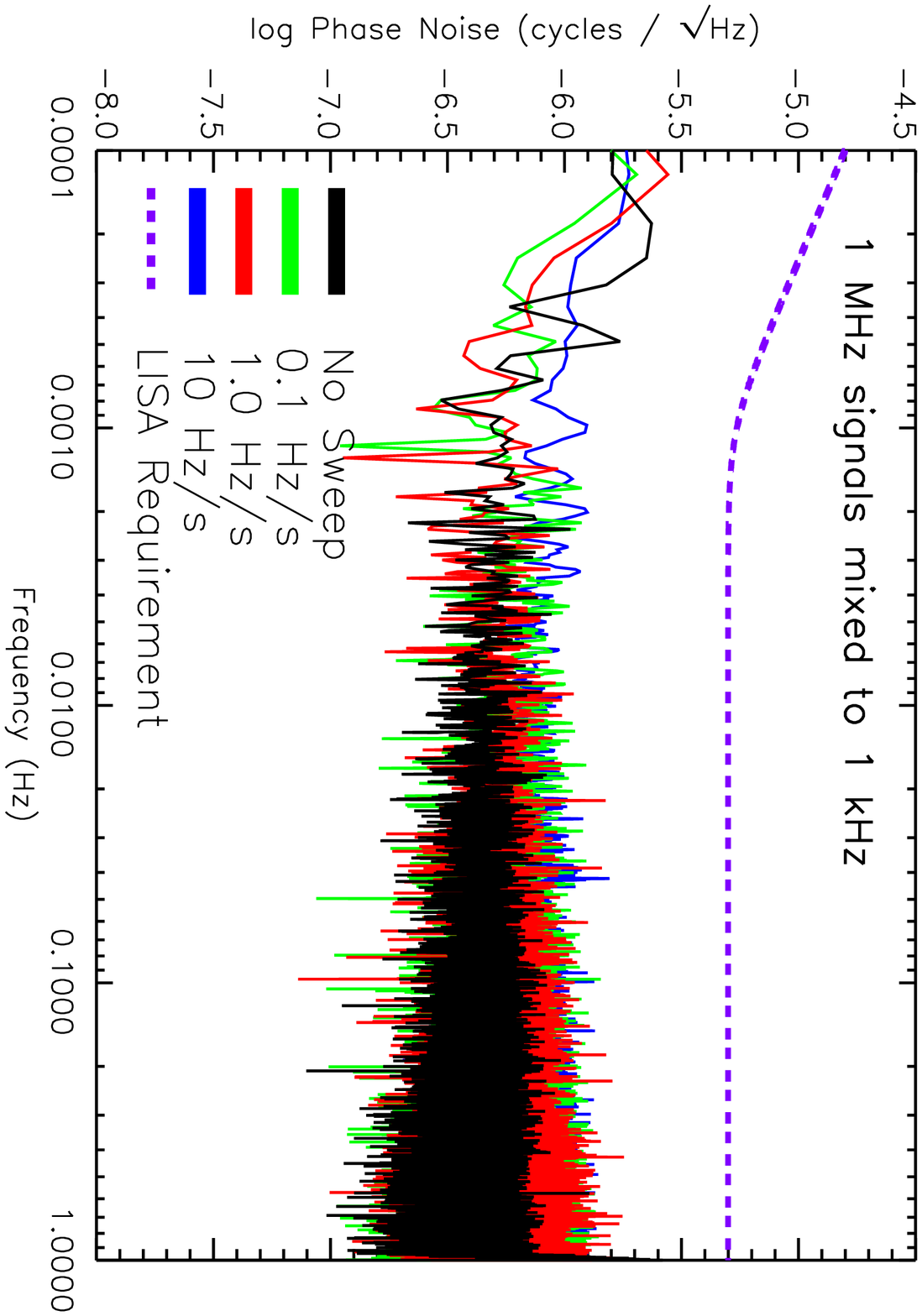}
\caption[Frequency Sweep Rate Investigation]{\label{fig:rate}
Phase noise spectra for 1\,MHz signals with various sweep rates mixed to 1\,kHz.
We found the phase noise level to be unchanged
for sweep rates as fast as 1\,kHz/sec.
}\end{idlplot}
\begin{idlplot}{!p}{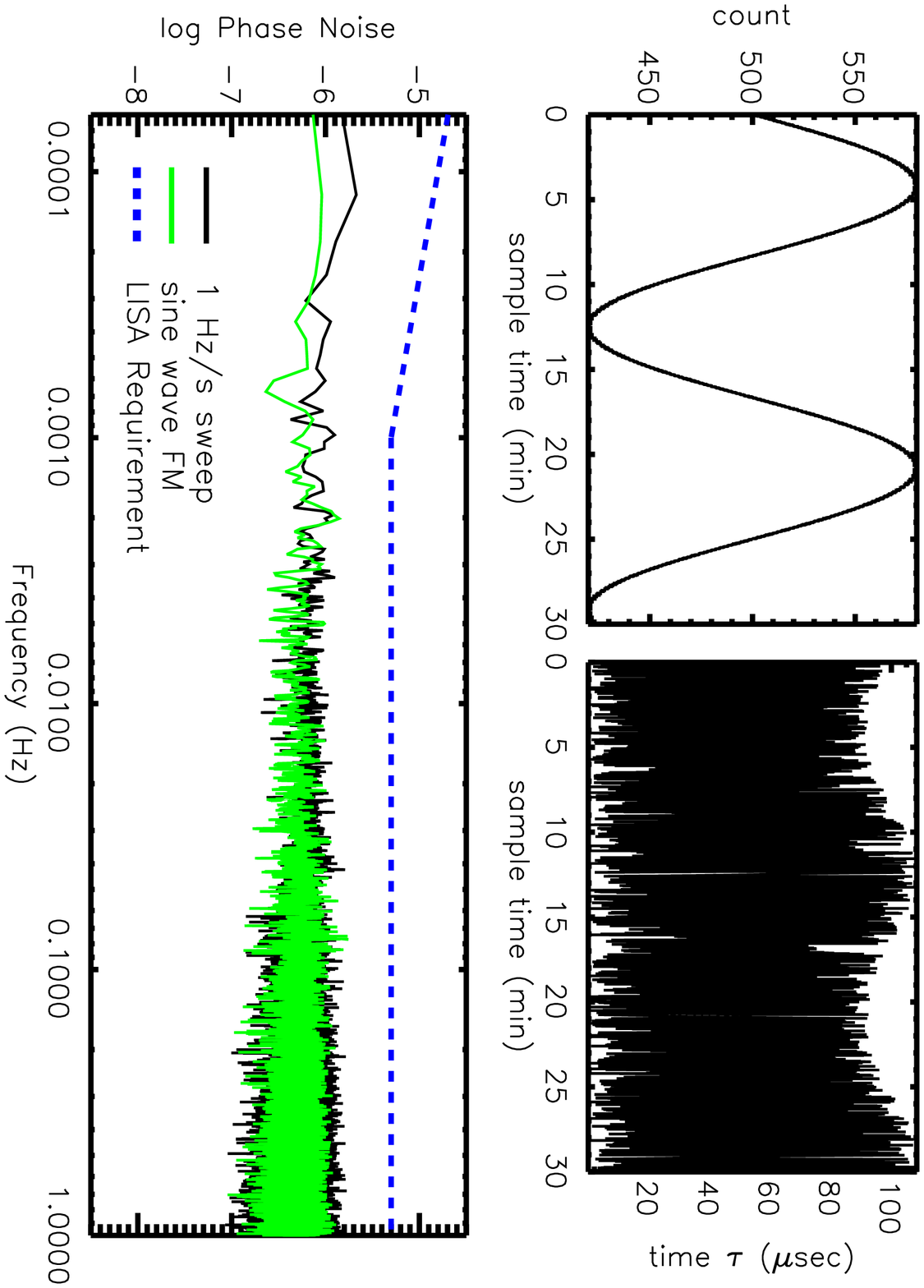}
\caption[Sweep Acceleration Investigation]{\label{fig:wave}
A 1\,MHz signal is frequency modulated by
a 160\,Hz depth 0.001\,Hz sinusoid.  The signal is mixed down to 1\,kHz for
phase measurement.   The sample frequency is 2\,Hz.
{\bf Top panels:} The counts and times follow the frequency modulation of the
signal as expected.
{\bf Bottom panel:} Phase noise of the data in the previous panels.
The ``1\,Hz/s sweep'' data is for a constantly swept signal at 1\,MHz.
Having the sweep rate change appears not to affect the phase noise
level in this situation.
}\end{idlplot}

Of possible concern are jumps in phase of the LO
during frequency steps.  Our LO has good phase continuity and
the jumps in phase are smaller than the resolution of our
phasemeter.  Feedthrough of these phase jumps to the IF signal
are not resolvable.

\subsection{Sweep Acceleration Investigation}

The distances between the sciencecraft 
can be approximated as sinusoids in time (see \cite{sphughes} for a
preliminary description of the LISA orbits).
The period of the sinusoid of the orbits is on the order of one year.  
For a 1\,Hz/s maximum sweep rate, the maximum sweep acceleration
will be about 0.1\,$\mu$Hz/s$^2$.  
We can simulate a changing sweep rate but only one with a much shorter period.  
The maximum modulation period of our frequency synthesizer is 1000~seconds.  
To attain a maximum sweep rate of 1\,Hz/s, we frequency 
modulate our source with a modulation depth of 160\,Hz.  
This yields a sweep acceleration of approximately 6.3\,mHz/s$^2$.
Figure~\ref{fig:wave} contains the results from this experiment which 
are as expected:
the counts follow a sinusoidal pattern with a mean of 500
(equating to a 1\,kHz signal sampled with a 2\,Hz clock)
and an amplitude of 80 (equating to a modulation depth of 160\,Hz).
The phase noise appears to be unchanged from that of a signal with
a linear frequency ramp at 1\,Hz/s.

\subsection{Bandwidth Investigation\label{bandwidth}}

The phase noise for a sweeping frequency
signal is dependent on the operating bandwidth of the phasemeter.  
In the majority of our data we have implemented a 1\,kHz operating bandwidth
for our phasemeter.  
Figure~\ref{fig:bandwidth} contains phase noise
results with various
operating bandwidths of our phasemeter between 10\,Hz and 1\,MHz.   
\begin{idlplot}{t}{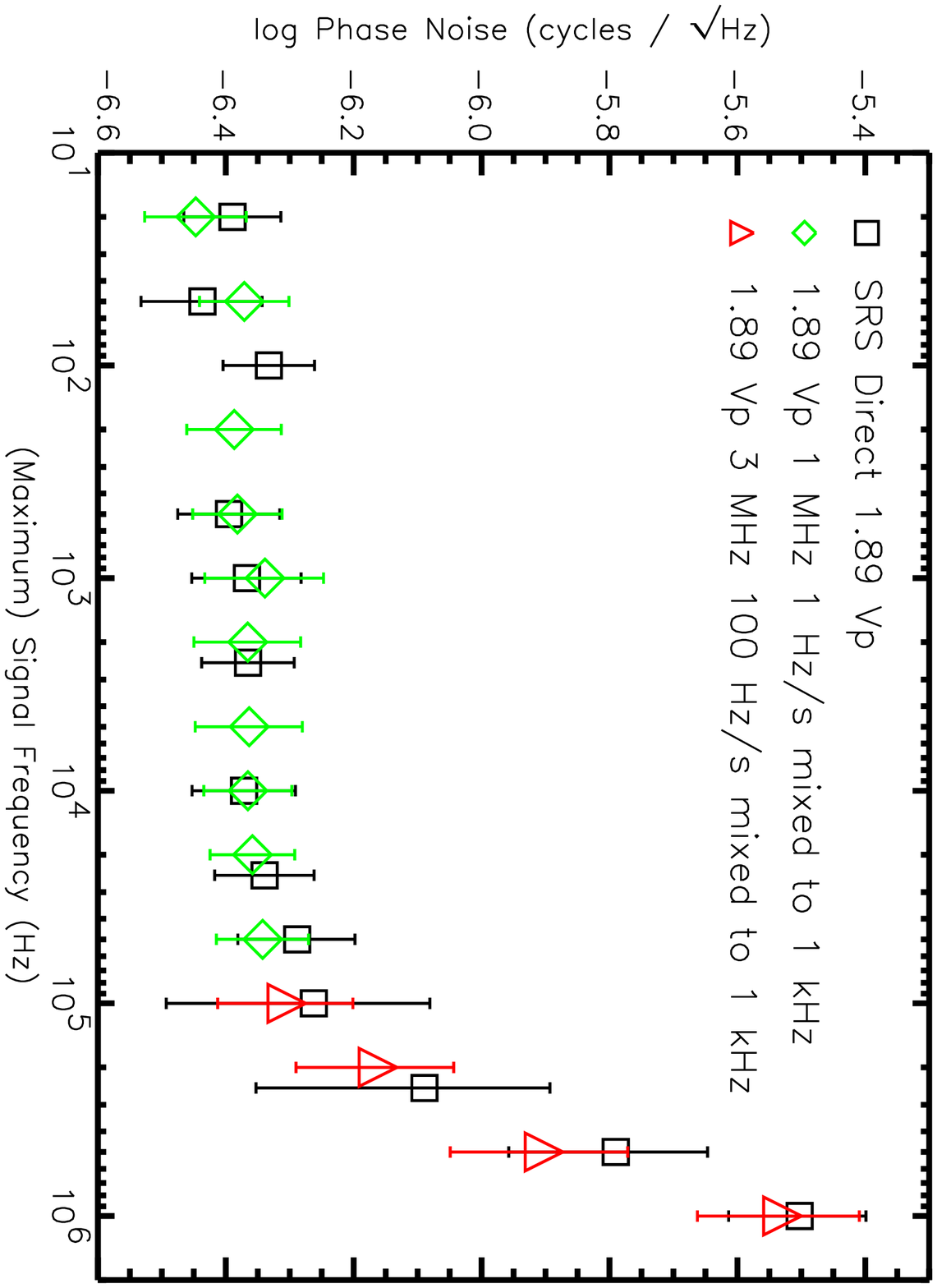}
\caption[Phase Noise vs Phasemeter Bandwidth]{\label{fig:bandwidth}
Phase noise data for various operating bandwidths of our phasemeter.
The ``SRS Direct'' data was taken without mixing the signal.
The phasemeter operating bandwidth when mixing follows
the phasemeter operating bandwidth without mixing.
}\end{idlplot}
A 1\,MHz signal swept at 1\,Hz/s was used
to investigate bandwidths smaller than 10\,kHz.  Above 10\,kHz
a 3\,MHz signal with a sweep rate of 100\,Hz/s was used.
The LO was instructed to step-up in frequency when the IF
reached the limit of the operating band.
It is apparent that a smaller bandwidth for the phasemeter
is advantageous for keeping the phase noise suppressed.  The phase
noise floor is the noise of our frequency synthesizer at about
$0.42\,\mcph$ above 1~mHz for a 1.89~volt-peak signal.  
As shown in the figure, the phase noise appears to follow
the phase noise of signals without mixing.
The small decrease in phase noise for maximum signal frequencies
above about 100\,kHz between the triangle and square data 
is due to the reduced amount of time the signal spent at the higher frequencies.
The maximum bandwidth which yields the
noise floor of our electronics is approximately 100\,kHz.  
Our choice of a 1 to 2\,kHz operating band for our phasemeter seems practical.

\section{A LISA Test-Bed Interferometer}

The LISA fringe signal has three important qualities: 
(1) a baseband fringe frequency which ranges from zero to tens of MHz, 
(2) the fringe sweeps at a rate up to 1\,Hz/s, and
(3) the fringe has a small signal level 
resulting from the heterodyning of 100\,pW
and 1\,mW laser beams.  Our table-top interferometer has
been designed to simulate these three aspects.  In addition, 
our table-top interferometer produces a fringe with a phase noise
below $5\,\mcph$ in the 1\,mHz to 1\,Hz band, which is the
part of the pathlength error budgeted to the phase measurement system \cite{PPA}.

\begin{sidewaysfigure}
\includegraphics[width=9.0in,angle=0]{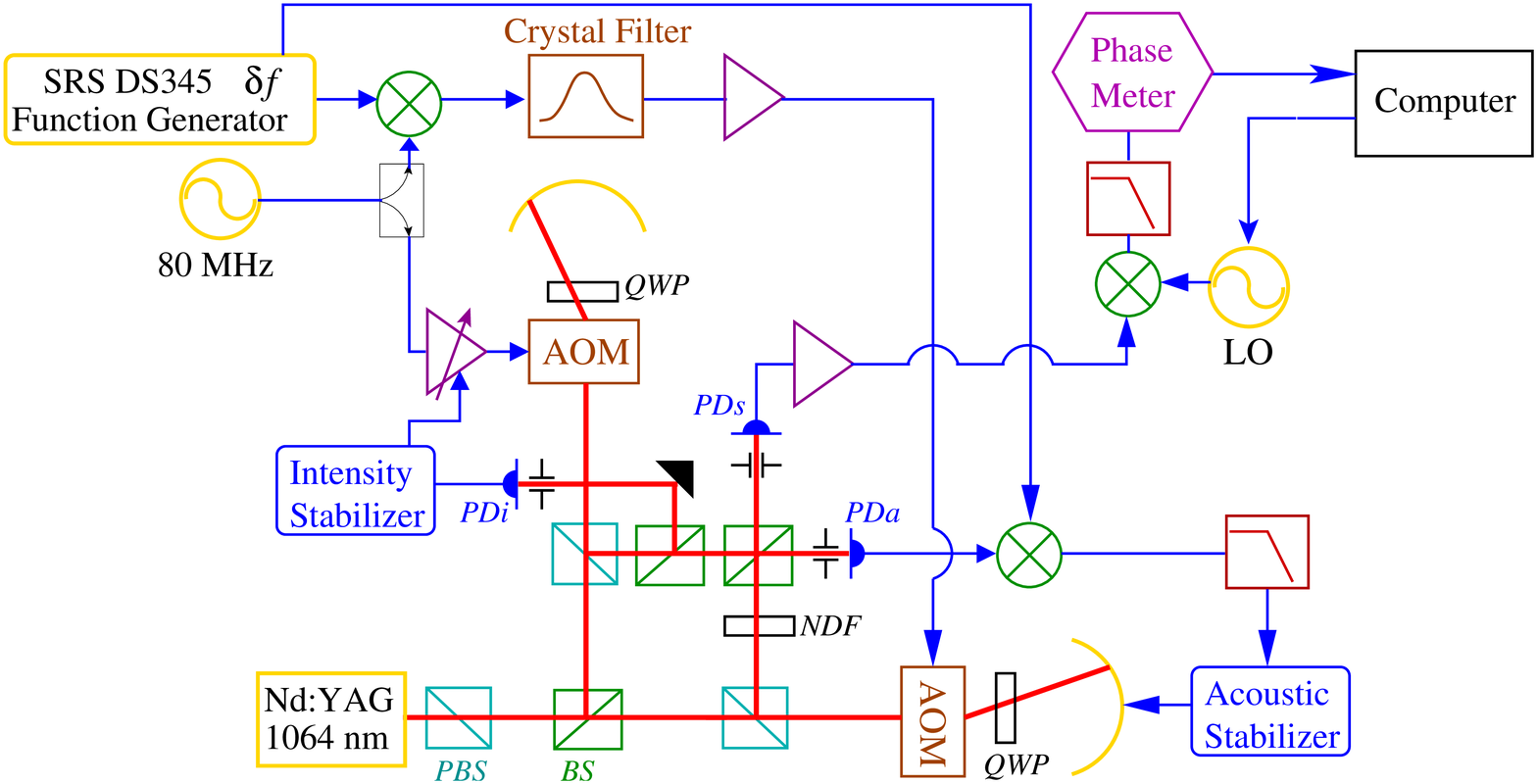}
\parbox{9in}{
\caption[Full Interferometer Layout]{\label{fig:ifmfull}
A schematic of our table-top fringe generator with acoustic stabilization, 
frequency generation, and phasemeter included.  
The general layout is that of Figure~1 in \cite{jennrich}.
The neutral density filter (NDF) reduces the power in
the horizontal arm to 100\,pW.
The intensity stabilization loop is described in the text.
}}\end{sidewaysfigure}

A detailed description of our experimental setup was given in \cite{jennrich}.
Figure~\ref{fig:ifmfull} contains an updated schematic of
our interferometer layout.
The labeling of components is that
of \cite{jennrich}.  Included in this schematic are
our acoustical stabilization loop described in \cite{jennrich}, 
our frequency generation technique described in \cite{pollack}, 
our data taking zero-crossing phasemeter with LO feedback,
and the new intensity stabilization loop
on the bright arm of the interferometer discussed in the following section.

\subsection{Constant Frequency Data}

One of the first steps in verifying our table-top 
interferometer is to demonstrate
the LISA phase measurement requirement of a constant frequency signal 
created by heterodyning a low power laser beam with a high power laser beam.  
We adjust the power differential between the arms
by using neutral density filters.
Although our commercial laser has a built-in intensity noise eater,
residual laser intensity noise, at frequencies below 1\,MHz,
in the bright arm of the interferometer
dominates the phase noise level for large power differentials (see
Figure~\ref{fig:powerlaw} below).
The relationship between the phase
noise and the power differential is easy to derive. 

The total electric field at the photoreceiver is the sum of the contributions
from each arm of the interferometer.  If we write the electric field in
each arm of the interferometer as 
$E (x, t) =  \sqrt{P}\, e^{\imag (k x - 2 \pi f t + \phi)}$
then the intensity measured at the photoreceiver will be
\begin{equation}
P_\mathrm{meas} = P_1 + P_2 + 2\, \sqrt{P_1 P_2}\, \cos\, ( 2 \pi\, \delta\!f\, t - \delta\phi)\,,
\end{equation}
\noindent where $\delta\!f = f_2 - f_1$ and $\delta\phi = \phi_2 - \phi_1$.  
The fringe signal is represented in the above equation by the 
sinusoidal cross term.  The other two terms are DC offsets.
The frequency of the fringe is naturally the difference
frequency between the two arms of the interferometer set by our
frequency synthesizer.  
The phase difference $\delta\phi$ can be adjusted by dithering the
spherical mirror at the end of one arm of the interferometer.

We write the power in the dim arm as $P_2 = \rho\, P$,
where $\rho$ is the power ratio, or inverse power differential,
and the power in the bright arm is $P_1 = P$.
With this definition the amplitude of the fringe becomes $P \sqrt{\rho}$,
assuming a 50-50 beamsplitter.
Therefore the amplitude of the fringe is inversely proportional to the 
square root of the power differential between the arms of the
interferometer.  
Since the phase noise level, $\phi_n$, is proportional to the fractional
error, we obtain
\begin{equation}\label{eqn:dp}
\phi_n\, (\cph) \propto \frac{V_n}{P \sqrt{\rho}}\,,
\end{equation}
\noindent where $V_n$ is the intrinsic noise level of our electronics.
We see in this relation that the phase noise level
is proportional to the square root of the power differential
which is consistent with the data shown in Figure~\ref{fig:powerlaw}.

Since the dim beam is reduced in power by several orders of magnitude,
the main contributor of amplitude noise in the final fringe signal is the
intensity noise of the bright beam.
Before the recombining beamsplitter the bright beam passes through
a ``pick-off'' beamsplitter.  This beamsplitter is actually a 50-50
beamsplitter so that half of the light in the bright beam is incident
upon a separate photoreceiver, PDi (see Figure~\ref{fig:ifmfull}), which is 
used for the stabilization of the intensity of the laser 
in the bright beam.  

The output of PDi goes to a servo controller
which attempts to remove variations in the intensity of the
received light.  Intensity control is accomplished by changing the
amplitude of the frequency input to the AOM in the bright arm.
This gain control is symbolized in the schematic of Figure~\ref{fig:ifmfull}
as an amplifier with an arrow through it.

\begin{idlplot}{t}{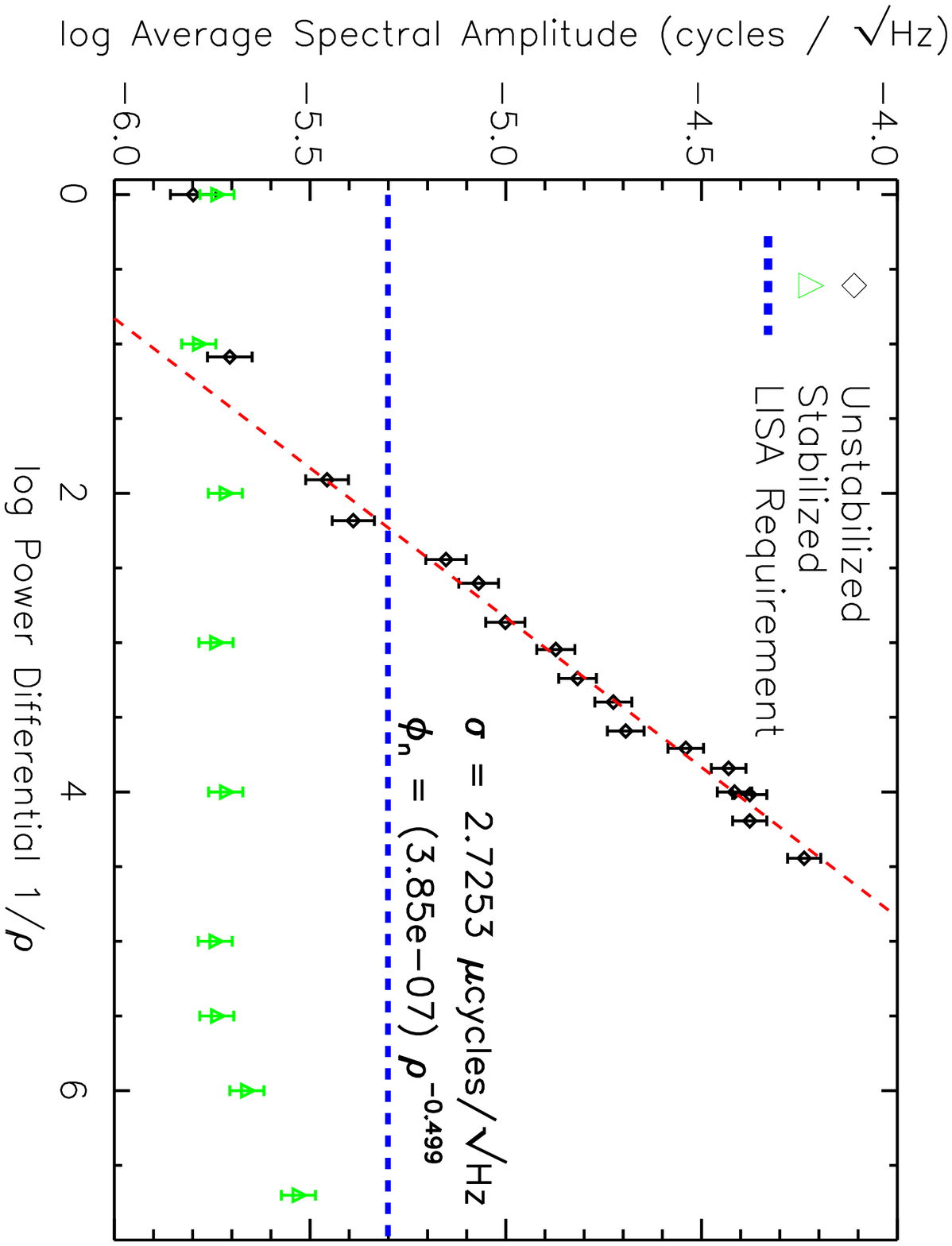}
\vspace*{-0.4cm}
\caption[Phase Noise Dependence on Power Differential]
{\label{fig:powerlaw}
Integrated phase noise in the 1--100\,mHz band for a range of power
differentials between the arms of the interferometer.  
Without intensity stabilization the phase noise rises as the square
root of the power differential as expected.  
}\end{idlplot}

Figure~\ref{fig:powerlaw} contains phase noise levels
of 50\,kHz fringe signals taken with various power differentials.  
With the intensity stabilization servo off the data follows 
the relation in Equation~(\ref{eqn:dp}).
With the intensity stabilization servo active the intensity
noise of the laser is no longer the dominant noise source
for power differentials greater than 10.  The intensity noise
resumes the role of being the largest noise source for power
differentials greater than about one million.

With the intensity stabilization servo active the power level in
the bright arm of the interferometer drops to 0.5\,mW in order
to facilitate the suppression of intensity noise.  The power in
the dim arm is unchanged from the target power level of 100\,pW.
These power levels create a power differential of five million.
The required power differential has been stated historically 
to be ten million.  This comes from the ratio of 1\,mW and 100\,pW.
There is no hard and fast requirement on there being 1\,mW of light
used from the local laser beam.  Therefore we feel
confident that 0.5\,mW is sufficient for our demonstration.

\subsection{Timing a LISA-like Fringe\label{s4:lisalike}}

We are now in position to produce a full LISA-like fringe from 
our table-top interferometer.  
We choose a collection of baseband frequencies ranging from
50\,kHz to 20\,MHz for data collection.  For each baseband
frequency the signal frequency slowly sweeps at a rate of 1\,Hz/s.
In addition, there is a power differential between
the arms of the interferometer of five million.
Phase noise spectra of the resulting data
are shown in Figure~\ref{fig:specsweep}.  
For this data we have instructed the LO to keep the IF between
1 and 2\,kHz.

\begin{idlplot}{b}{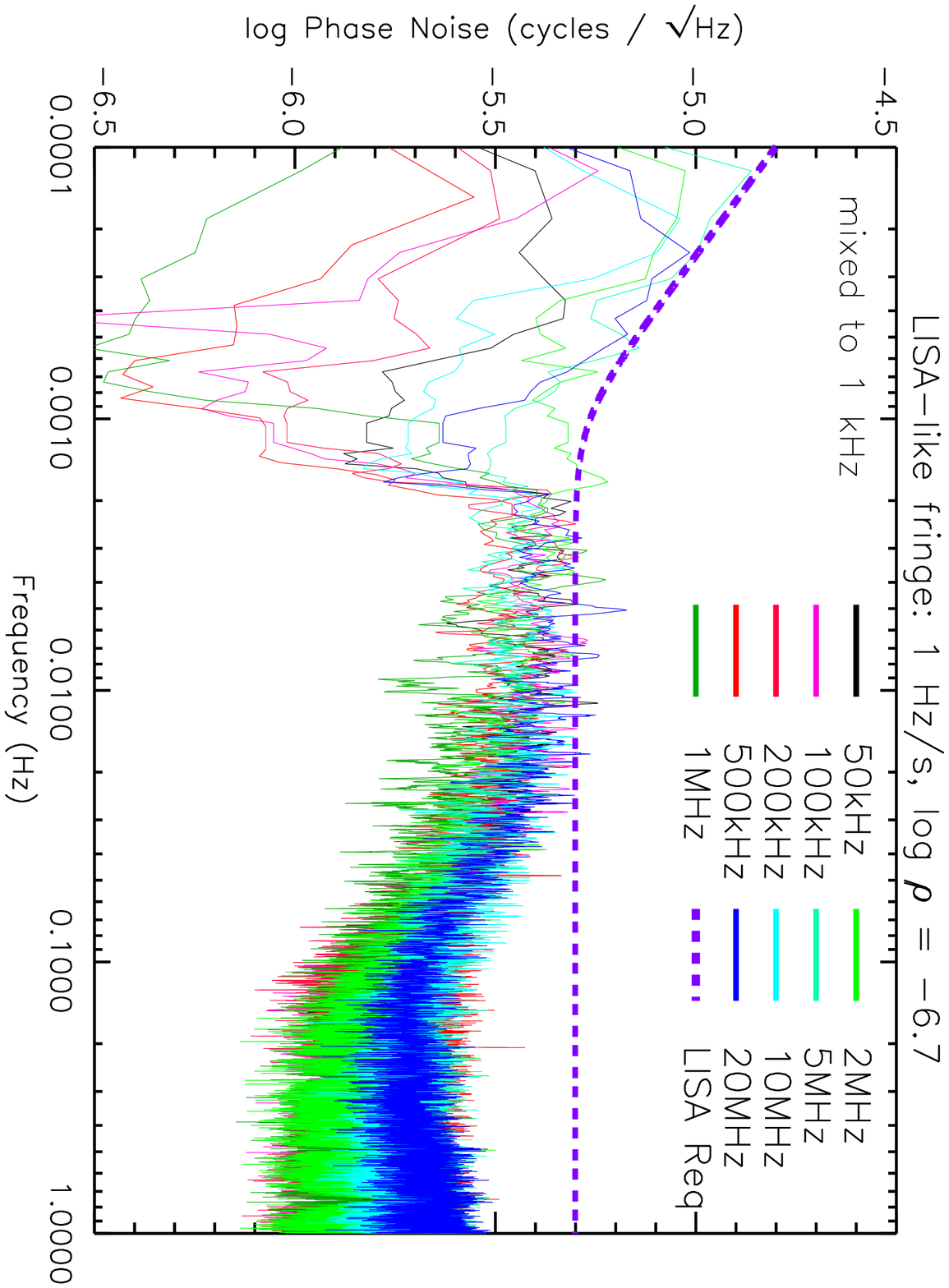}
\caption[Phase Noise of LISA-like Fringes]{\label{fig:specsweep}
Phase noise spectra of LISA-like fringes at a range of baseband
frequencies.  
The LO was instructed to keep the IF between 1 and 2\,kHz.  
The power levels of the arms of the interferometer
are 0.5\,mW and 100\,pW, resulting in a power differential of five
million.  The fringe sweep rate is 1\,Hz/s.  
}\end{idlplot}

All of the spectra shown in Figure~\ref{fig:specsweep} have the
same general phase noise profiles at frequencies above 2\,mHz
from the fringe frequency.
With the intensity stabilization servo active the phase noise
is dominated by electronic noise from the frequency generation technique.
We are meeting the LISA requirement of 5\,$\mcph$ 
between 1\,mHz and 1\,Hz for all baseband frequencies examined.
Also, the noise only rises as roughly $f^{-1/2}$ below 1\,mHz,
which is well below the assumed increase of $f^{-2}$ that
would be allowed for LISA.

It is interesting to note that in general there appears to be
less phase noise at frequencies below 2\,mHz from the fringe frequency
for low frequency fringes.
We are unsure of the cause of this, and have not been able to
reproduce these low-frequency stable phase signals.
Our subsequently taken data always rises as $f^{-1/2}$ at low frequencies.
Note that 
the data presented are each the average of four 4.5-hour long data sets.  
At 0.1\,mHz there are 1.6 cycles in 4.5 hours, and 
consequently there are 6.6 cycles in each spectra of 
Figure~\ref{fig:specsweep}.
In contrast for a 1\,mHz signal there are 66 cycles in each
spectra.  The smaller number of cycles might lead one to expect that
the error at 0.1\,mHz would be 3 times larger than the error at 1\,mHz.
In actuality, the variance of each power spectrum, at all frequencies,
is roughly the mean power level.
For our spectra the mean power level is about 2\,$\mcph$
integrated over the $10^{-4}$ to 1\,Hz band.  
This error should be kept in mind when considering
the suspiciously low phase noise values below 2\,mHz.

\subsection{Sweeping the Local Oscillator}

In Section~\ref{bandwidth} we examined cases
where we allowed the IF to roam over the full practical range of the phasemeter
operating band as well as confining the IF to a small bandwidth.
We can take that investigation one step further and have the LO continuously
sweep in frequency.  
Phase delays through the crystal filter used in our frequency
generation technique (described in Section~3.2 of \cite{pollack})
very slightly alter the actual sweep rate of the fringe signal.
Since the sweep rate is not exactly equal
to the sweep rate programmed into our frequency synthesizer,
we feedback to the LO during data taking, changing the sweep rate
until it matches the sweep rate of the fringe signal.
In this way the IF remains constant in frequency.
There is a similar situation in LISA where the sweep rates of the fringe signals
may not be known exactly.  If LO tracking is to be used, 
then a similar feedback loop to ours will need to be employed.
We found that the final phase noise was 
unchanged from the case of a constant fringe frequency.

\section{Items of Concern\label{problems}}

There are several aspects of our experimental simulation which
are different from LISA.  In this section we discuss the following
items of concern:
(1) correlation between the photoreceivers PDs and PDa at
the combining beamsplitter,
(2) laser frequency noise,
(3) aliasing of noise due to our analog demodulation and
decimation of sampling,
and (4) effects of tones on the LISA fringe.

\subsection{Correlation of Output Beams}

Observation of Figure~\ref{fig:ifmfull} indicates 
that the signal observed on PDs should be 
quite similar
to $\delta\!f$ due to the acoustical stabilization
feedback loop.  If the two signals were identical, then the
inclusion of a table-top interferometer in our
experiment serves no practical purpose.
However, this is not the case. 
We have measured the correlation between PDa and PDs to be 80\%
at 1\,mHz, falling rapidly to 0\% by 7\,Hz away from $\delta\!f$.



The correlation between PDs and PDa not withstanding, our
interferometer does provide some aspects of realism.
In practice we are looking for noise sources which are
not common-mode between PDs and PDa.  Naturally
these noise sources, and the methods we have
developed to deal with them, may be of importance
for LISA.  In particular,
the shot noise and electronic noise in each of our
photoreceivers is not common-mode and will appear 
in our measurement.

In addition, we use a realistic laser (although we ignore
laser frequency noise, see below), with a power differential
between the arms of the interferometer, and
a fringe beatnote ranging in frequency from nearly DC
to 20\,MHz, sweeping at a rate of 1\,Hz/s.
Residual laser amplitude noise is present in our experimental
setup and is a dominant noise source for large power
differentials between the arms of the interferometer.
A technique for suppressing amplitude noise will be necessary
on LISA, although a technique similar to ours will most likely not
be used.  However, it is probable that for ground testing of 
the LISA sciencecraft a technique similar to ours may be utilized.

\subsection{Laser Frequency Noise}

Laser frequency noise is essentially absent from our experiment
since we use only one laser.  However, we can simulate
the effect of laser frequency noise by injecting noise into
our interferometer by using a noisy $\delta\!f$ from
our SRS DS345 frequency synthesizer.
Any noise on this signal appears on our photoreceiver
and will not be removed by our acoustic stabilization loop.
We have done this experiment, and the final phase noise
level matches that expected due to the noisy source.
Of course our assumption in our result is that laser
frequency noise adds linearly and therefore can
be removed by this technique to the required precision from time-delay
interferometry (TDI, see e.g., \cite{TEA, TSSA} and references
therein).

\subsection{Aliasing of Noise\label{noise}}

There are two separate issues regarding aliasing.
In particular, the aliasing of laser frequency
noise is of concern.
The first issue is due to decimation of sampling.
If we sampled at the same frequency
as the IF then we would obtain all the
phase information possible.  However, since
we sample the phase at our clock frequency,
which is considerably less than the IF,
one might expect that we are losing phase information.
However, we count the zero-crossings between timing
measurements and therefore do not actually lose
any phase information.  
In effect we average the phase over the number of cycles
between clock ticks.

The issue of aliasing regarding decimation is that 
signals of frequencies higher than our clock frequency
will be aliased into our measurement band.  Indeed,
we see this effect in simple experiments, such as the following:
a 1\,kHz sinusoid is phase modulated with a $10^\circ$ 11\,Hz
signal.  Our clock frequency is set to 10\,Hz.  The resulting
phase noise spectrum shows a peak at 1\,Hz, which is the 11\,Hz
signal aliased by 10\,Hz.  
However, uncorrelated signals which alias into our measurement
band will be suppressed.  
This is because we are effectively averaging between clock
ticks as mentioned above.  
In this manner the noise at frequencies above our clock frequency 
aliased into our measurement band do not adversely effect
the measured phase noise level.

The second aliasing issue is due to the 
analog demodulation performed
to bring the frequency $\delta\!f$ to a suitable
IF for our phasemeter.  Laser frequency noise
at frequencies much larger than our IF away
from the carrier prior to demixing will appear 
at negative frequencies afterward.  Since our
phasemeter only detects positive frequencies,
this DC-wrapped noise will adversely affect
the measured phase.

The issue of DC-wrapping aliasing is not a deterrent for
the zero-crossing phasemeter.  Time-delay interferometry 
(TDI, see e.g., \cite{TEA, TSSA} and references therein)
is the process by which laser frequency noise will be
corrected for in LISA.  In TDI two phasemeter outputs
are combined with a time offset in order
to subtract the laser frequency noise contribution
to the phase measurement.

Naive thinking might lead one to believe that
aliased laser frequency noise would spell the end for TDI.
However, even though noise is being aliased
into the measurement band, it is being
aliased in the same way in each measurement.
Therefore the construction of TDI variables
will still correct laser frequency noise,
even though that noise has been aliased.

We have performed TDI-like experiments with our
table-top interferometer and zero-crossing phasemeter.
In the most complex of these experiments we
inject a noisy signal into our interferometer.
This signal is mixed down to a suitable IF
for our phasemeter.  In this process we would
expect that aliased noise from the analog demodulation
of the high frequency fringe signal 
will dominate the phase noise measurement.  Indeed it does.
We take two such data sets.  The noise is programmed,
therefore it is identical in the second data set.  However,
due to slight timing errors in the computer GPIB communication
process the noise will be time-delayed.
In software we fit for the time-delay between the
two data sets.  After subtraction the noise level
is $\sqrt{2}$ higher than the un-noisy signal
from the interferometer as expected.

It should be mentioned that we have relied on the response
of our phasemeter to be nearly identical during the two data sets.
The precision of our phasemeter is crucial in allowing us
to subtract the time-delayed noisy signals to below the
shot noise level of our interferometer.
A slightly more advanced test of this TDI-like experiment
would be with two separate phasemeters measuring
simultaneously.  This would be possible in our
current setup if we added a phasemeter at the PDa port
of our interferometer.

We have determined that aliased laser frequency noise can
be successfully canceled.  
Therefore
the most serious aliasing issues discussed above
are not deterrents from using the zero-crossing
phasemeter in LISA.
However, shot noise or technical noise that is
DC-wrapped aliased in would not be corrected for in the TDI process,
and therefore the noise level will be $\sqrt{2}$ higher than
the shot noise alone, just as in our demonstration.

\subsection{Communication Tones}

Our table-top interferometer can be easily modified
to include USO sidebands, ranging tones, and
digital data modulations on the laser beam.
We have performed the necessary modifications and
will be presenting our results in a 
another article \cite{comm}.

\section{Summary}

We have constructed a table-top interferometer which produces
LISA-like fringes.  These fringes have the following properties:
(1) a baseband fringe frequency which ranges from 50\,kHz to 15\,MHz,
(2) the fringe frequency sweeps at a rate of 1\,Hz/s,
and (3) a small signal level resulting from the heterodyning of a
100\,pW laser beam and a 0.5\,mW laser beam.

We have verified the power differential-phase noise relation
Equation~(\ref{eqn:dp}) and have reduced this dependence by the addition
of an intensity stabilization servo for the bright arm of the
interferometer (Figures~\ref{fig:ifmfull} and \ref{fig:powerlaw}).

We have demonstrated that the fringes produced by our fringe generator
have a displacement noise less than the LISA
allocation to phase measurement, namely 
5\,pm/$\sqhz$ from 
1\,mHz to 1\,Hz, rising slower than $f^{-1/2}$ at lower frequencies
(Figure~\ref{fig:specsweep}).

We have investigated the effects of tracking the slowly varying fringe
with the local oscillator to produce a constant frequency signal
for our phasemeter.  As will be the case for LISA we found that small
corrections to the LO sweep rate were needed to properly account for the
changing fringe rate.  Upon implementing these modifications we were able to
attain a constant frequency IF which had a phase noise level identical
to the case of a non-sweeping fringe signal.

\ack

Support for this work has been provided under NGT5-50451 and S-73625-G.  
We wish to thank Peter L. Bender, John Hall, and Jun Ye for their
generous insight to aspects of this project.
We would also like to thank Daniel A. Shaddock for several
stimulating conversations relating to the zero-crossing phasemeter.

\end{document}